\documentstyle[epsf,onecolumn]{mn}

\def\etal{et al. }  \def\eg{e.g.} 
\def\lap{\hbox{${_{\displaystyle<}\atop^{\displaystyle\sim}}$}}

\newif\ifAMStwofonts
\AMStwofontstrue

\title{Simulations of Glitches in Isolated Pulsars}

\author[M.B. Larson and B. Link]
       {Michelle B. Larson and Bennett Link \\
        Department of Physics, Montana State University, Bozeman, MT. 
        59717}

     \date{Accepted 2001 xxxx xx
      Received 2001 xxxx xx;
      in original form 2001 xxxx xx}

\pagerange{\pageref{firstpage}--\pageref{lastpage}}
\pubyear{2001}
\begin{document}


\topmargin = -2pc
\maketitle
\label{firstpage}

\begin{abstract}
Many radio pulsars exhibit glitches wherein the star's spin rate 
increases fractionally by $\sim 10^{-10} - 10^{-6}$.  Glitches are 
ascribed to variable coupling between the neutron star crust and its 
superfluid interior.  With the aim of distinguishing among different 
theoretical explanations for the glitch phenomenon, we study the 
response of a neutron star to two types of perturbations to the vortex 
array that exists in the superfluid interior: 1) thermal motion of 
vortices pinned to inner crust nuclei, initiated by sudden heating of 
the crust, (e.g., a starquake), and 2) mechanical motion of vortices, 
(e.g., from crust cracking by superfluid stresses).  Both mechanisms 
produce acceptable fits to glitch observations in four pulsars, with 
the exception of the 1989 glitch in the Crab pulsar, which is best fit 
by the thermal excitation model.  The two models make different 
predictions for the generation of internal heat and subsequent 
enhancement of surface emission.  The mechanical glitch model predicts 
a negligible temperature increase.  For a pure and highly-conductive 
crust, the thermal glitch model predicts a surface temperature 
increase of as much as $\sim$ 2\%, occurring several weeks after the 
glitch.  If the thermal conductivity of the crust is lowered by a high 
concentration of impurities, however, the surface temperature 
increases by $\sim$ 10\% about a decade after a thermal glitch.  A 
thermal glitch in an impure crust is consistent with the surface 
emission limits following the January 2000 glitch in the Vela pulsar.  
Future surface emission measurements coordinated with radio 
observations will constrain glitch mechanisms and the conductivity of 
the crust.

\end{abstract}

\begin{keywords}

stars: interiors --- stars: neutron --- stars: evolution --- 
stars: rotation --- superfluid --- dense matter
\end{keywords}

\section{Introduction}

Many pulsars exhibit glitches, sudden jumps in spin rate, superimposed
on the gradual spin down due to electromagnetic torque (see, \eg,
Lyne, Shemar \& Smith 2000).  Glitches involve fractional jumps in
spin rate of $\Delta\nu/\nu\simeq 10^{-10}$ to $10^{-6}$, with
recovery to the pre-glitch spin-down rate occurring over days to
months in most cases.  Some pulsars show no obvious recovery, and
continue to spin down faster than had the glitch not occurred.  The
1989 glitch of the Crab pulsar ($\Delta \nu/\nu \simeq 7\times
10^{-8}$) was partially time-resolved \cite{Lyne}.  This glitch showed
a quick rise on a timescale of hours with additional spin-up taking
place over approximately one day.  In contrast, the Vela ``Christmas''
glitch ($\Delta \nu/\nu \simeq 2\times 10^{-6}$) observed in December
of 1988 \cite{Vela} showed much different behaviour.  In this case the
glitch was not time-resolved, and occurred in under two minutes.  The
January 2000 glitch in the Vela pulsar ($\Delta \nu/\nu \simeq 3\times
10^{-6}$) was similar to the Christmas glitch \cite{DMC}.  A
number of pulsars (the Crab in particular) exhibit permanent increases
in spin-down rate after a glitch occurs, typically
$\Delta\dot\Omega/\dot\Omega \simeq 10^{-4}$.  In the Crab, these
offsets produce much larger cumulative timing residuals than the
glitches themselves.  In addition to glitches, nearly all pulsars
exhibit low level fluctuations in their spin rate, timing noise,
believed to be of a different origin than glitches (see \eg,
D'Alessandro \etal 1995).

Glitches are thought to represent variable coupling between the
stellar crust and the superfluid interior.  Two questions concerning
the glitch phenomena are: 1) where in the star the coupling occurs,
and, 2) how the coupling is triggered.  The rotation of the neutron
superfluid interior is governed by the dynamics of vortex lines; a
spin jump of the crust would result from sudden motion of vortices
away from the rotation axis.  In the inner crust, vortices could {\em
pin} to the lattice \cite{AI,Alp77,ALS,EB88}, allowing the superfluid
to store angular momentum as the crust spins down under
electromagnetic torque.  As a velocity difference between the solid
and the superfluid develops, vortices {\em creep} through the crust at
a rate that is highly sensitive to temperature
\cite{Alp77,ACP,LE91,LEB,CC93a,CC93b}.  Based on an idea by
Greenstein (1979a,b), Link \& Epstein \shortcite{LE} have proposed a
{\em thermal glitch} mechanism in which a temperature perturbation
causes a large increase in the vortex creep rate; in consequence,
the superfluid quickly loses angular momentum and delivers a spin-up
torque to the crust. A candidate mechanism for providing the required
heat is a starquake arising from relaxation of crustal strain as the
star spins down. Starquakes could deposit as much as $\sim 10^{42}$
ergs of heat in the crust \cite{BP,Cheng92}. Ruderman
\shortcite{Rud91} has proposed a different model in which vortices
strongly pinned to the inner crust lattice stress the crust to the
point of fracture, resulting in outward motion of vortices with plates
of matter to which they are pinned.  In the core, pinning may occur
between the vortices and flux tubes associated with the
superconducting proton fluid \cite {CCD}, allowing the core
superfluid, or a portion of it, to store angular momentum.  Ruderman,
Zhu \& Chen (1998) have proposed a core-driven glitch mechanism in
which the expanding vortex array of the core forces the magnetic flux
into the highly-conductive crust, stressing it to fracture.  In this
model, crust cracking allows the core vortex array to suddenly
expand outward, spinning down a portion of the core superfluid and
spinning up the crust.  Carter, Langlois \& Sedrakian \shortcite{Car}
have suggested that centrifugal buoyancy forces are the origin of
pressure gradients sufficient to crack the crust, allowing outward
vortex motion.  Other proposed glitch mechanisms include catastrophic
unpinning of vortices in the crust \cite{Cheng88,AP,MI}, and vortex
motion at the crust-core boundary due to proton flux tube annihilation
there \cite{SC}.  In any of these crust or core-driven glitch models,
dissipation that accompanies outward vortex motion generates heat that
might produce detectable emission as the heat arrives at the stellar
surface.

Quantitative calculations are necessary to distinguish among different
models for the glitch phenomenon.  The thermal glitch model of Link \&
Epstein \shortcite{LE} produced good qualitative fits to glitch
observations in the Crab and Vela pulsars.  This paper is an extension
of that work with more realistic physical inputs and detailed modeling
of the timing data.  We include nonlinear thermal diffusion which has
the effect of slowing the glitch spin-up.  We also include the effects
of superfluid heating due to differential rotation between the
superfluid and the crust
\cite{Green,HGG,Alp87,SL,VR91,VREM,Umeda,VRLE,LL} and study the
propagation of heat to the stellar surface.  We consider two types of
rearrangement of the superfluid vortices: 1) thermal excitation of
vortices over their pinning barriers (a thermal glitch) and 2)
mechanical motion of vortices (a mechanical glitch).  The first case
models the response to sudden heating of the crust, \eg, from a
starquake.  The second case models catastrophic unpinning events,
vortex motion as a result of crust cracking due to superfluid
stresses, or core-driven glitches involving vortex motion near the
crust-core boundary.  In addition to simulations of the rotational
dynamics, we predict the characteristics of the emerging thermal
wave which could, in some cases, be visible from the surface of the
neutron star weeks to years after a glitch occurs.

This paper is organized as follows. In Section 2 we provide an 
overview of the physical setting and discuss the treatment of the 
coupled rotational and thermal dynamics we use in our simulations of 
pulsar glitches.  In Section 3 we discuss the details of our numerical 
models.  In Section 4 we present our simulations of the spin-up 
process and the emergence of the thermal wave at the stellar surface.  
We compare our simulations with spin observations of four pulsars and 
surface emission data following a recent glitch in the Vela pulsar.  
In Section 5 we conclude with discussion.

\section {Input Physics}
\label{input}

A neutron star consists of an atmosphere, an outer and inner crust, 
and the core.  The inner crust begins at a density of $\sim 4 \times 
10^{11}$ g cm$^{-3}$ and extends to approximately nuclear saturation 
density, $\rho_{o} = 2.8 \times 10^{14}$ g cm$^{-3}$.  In this region, 
a neutron-rich lattice coexists with a neutron superfluid, protons and 
relativistically degenerate electrons.  The inner crust lattice 
dissolves near nuclear density \cite{Lorenz,Pethick}.  Most of the 
mass of the core is expected to reside in superfluid neutrons and 
superconducting protons, with electrons and muons also present.

Glitch models that rely on the superfluid interior as an angular
momentum reservoir require a metastability of the vortex state to
sustain differential rotation between the solid and liquid components
of the star.  In the inner crust, the metastability could arise
through the pinning of vortices to the lattice.  The details of
pinning are uncertain.  The form of the vortex-nucleus interaction
potential is not well known due, in part, to uncertainties in the
nucleon-nucleon interactions and the structure of the vortex core.
Preliminary calculations of the vortex-nucleus interaction gave
energies of $\sim$ 1-10 MeV \cite{EB88}.  Pizzochero, Viverit \&
Broglia \shortcite{Pizz} refined these calculations and found
interaction energies on the order of several MeV per nucleus near
nuclear density.  By studying tension effects on the orientation of a
vortex line relative to crystal axes, Jones
\shortcite{Jon97,Jon98,Jon99} has argued that there can be significant
cancellation of the pinning forces on a vortex line.  The magnitude of
the cancellation is determined by the extent to which a vortex line
can bend to intersect pinning nuclei.  Link \& Cutler (2001) have
studied pinning of a vortex with finite tension, and estimate an
effective pinning force per nucleus of $\sim 10$ keV. Given the
uncertain nature of vortex pinning, we take the effective pinning
energy per nucleus to be a free parameter.  Our simulations fit the
spin data for effective pinning strengths of $\simeq 20-500 \ {\rm
keV/nucleus}$.

\subsection{Rotational Dynamics}
\label{RD}

The total angular momentum ${\mathbf J}$ of the star is that of the 
{\it effective crust} (the crust and all components strongly coupled 
to it) plus the angular momentum of the superfluid,

\begin{equation}
 {\mathbf J}_{\rm tot}(t) = I_c{\mathbf \Omega}_c(t) + \int dI_s 
 {\mathbf \Omega}_s({\mathbf r},t) = {\mathbf J}_0 - {\mathbf 
 N}_{{\rm ext}}t,
\label{ANG}
\end{equation}
where $I_{c}$ is the moment of inertia of the effective crust, $I_{s}$ 
is the crust superfluid moment of inertia, $I \equiv I_{c} + I_{s}$ is 
the total moment of inertia, ${\mathbf \Omega}_{c}$ is the angular 
velocity of the effective crust and ${\mathbf \Omega}_{s}$ is the 
angular velocity of the superfluid.  The initial angular momentum of 
the star is ${\mathbf J_{0}}$.  The star slows under an external 
torque, ${\mathbf N}_{{\rm ext}} \equiv I{\mathbf 
\dot\Omega}_{\infty}(t)$.  In rotational equilibrium the effective 
crust and the superfluid would spin down at the same rate, 
$\dot{\mathbf\Omega}_{\infty}(t)$.  The stellar core is thought to 
couple to the crust on timescales of less than a minute 
\cite{ALS,Abn}.  We therefore take the effective crust to include the 
mass of the core plus the crust, and the superfluid to exist between 
neutron drip ($\rho = 4.3 \times 10^{11}$ g cm$^{-3}$) and nuclear 
density.

We assume a geometry in which the angular velocity of the superfluid 
and crust are aligned with the external torque, as this is the state 
of lowest rotational energy for given angular momentum.  The rotation 
rate of the inner crust superfluid is determined by the arrangement of 
the vortex lines which thread it.  The equation of motion for the 
superfluid is (see e.g., Alpar \etal 1981; Link, Epstein \& Baym 
1993),

\begin{equation}
	{\mathbf \dot\Omega}_s({\mathbf r},t) = 
	-v_{cr}\left(\frac{2}{r_p} + \frac{\partial} {\partial 
	r_p}\right){\mathbf \Omega}_s({\mathbf r},t),
\label{SFD}
\end{equation}
where $r_{p}$ is the distance from the rotation axis and $v_{cr}$ is
the radial component of the average vortex velocity.  If vortex
pinning is relatively effective, as we assume, vortices can move
slowly with respect to the rotation axis through thermal excitations
or quantum tunnelling in a process of {\it vortex creep}.

The average velocity of vortex creep is determined by the 
vortex-nucleus interaction, the vortex core structure, the 
characteristic energy of excitations on a pinned vortex, and the 
velocity difference between a pinned vortex and the superfluid flowing 
past it.  Link \& Epstein \shortcite{LE91} and Link, Epstein \& Baym 
\shortcite{LEB} account for quantum effects and the vortex self-energy 
and obtain a creep velocity of the general form (see eq.  [6.9], Link, 
Epstein \& Baym 1993),

\begin{equation}
	v_{cr} = v_0 \exp (- E_{a}/T_{\rm eff}),
\label{creep}
\end{equation}
where $E_{a}$ is the activation energy for a pinned vortex segment to 
unpin.  The effective temperature is $T_{{\rm eff}} \equiv T_{q} {\rm 
coth} \frac{T_{q}}{T}$, where $T$ is the thermodynamic temperature and 
$T_{q}$ is the crossover temperature which determines the transition 
from vortex motion through thermal activation to that by quantum 
tunnelling \cite{LEB}.  The crossover temperature depends on the 
ground-state excitation energy of a pinned vortex.  In our simulations 
the stellar temperatures are much greater than the crossover 
temperature, so that thermal activation (classical creep) is the 
dominant creep mechanism.  In this limit $T_{{\rm eff}}$ reduces to 
the temperature $T$.  The multiplicative factor $v_{0}$ is comparable 
to the radial component of the velocity of an unpinned vortex line; 
we take its value to be $10^{6} \rm{cm \ s^{-1}}$ 
\cite{LE91,EB92,LEB}.

The mechanics and energetics of unpinning are affected by the vortex 
self energy, or tension, $\widehat{T}$.  For a vortex line with a 
sinusoidal perturbation of wavenumber $k$, the tension takes the form 
(Fetter 1967; see [Appendix A], Link \& Epstein 1991),

\begin{equation}
\widehat{T} = {\rho_{s}\kappa^{2}\over 4\pi}\Lambda,
\label{tension}
\end{equation}
where $\Lambda \simeq (0.116 - {\rm ln}k\xi)$ and $\xi$ is the vortex
coherence length.  The circulation associated with each vortex line is
$\kappa = h/2m_{n}$, where $m_{n}$ is the mass of a neutron and $h$ is
Planck's constant.  Typically $2 \leq \Lambda \leq 10$ in the inner
crust \cite{LE91}.  The relative importance of tension is determined
by the value of the stiffness parameter $\tau \equiv
\widehat{T}r_{o}/F_{p}\bar l$, where $F_{p}$ is the maximum
attractive force between a nucleus and a vortex, $r_{o}$ is the
range of the pinning potential, and $\bar l$ is the internuclear
spacing \cite{LE91}.  We take $F_{p} = U_{o}/r_{o}$, where
$U_{o}$ is the effective pinning energy per nucleus.  In terms of
fiducial values the stiffness $\tau$ is,

\begin{equation}
	\tau \simeq 
	100 \left({\rho_{s}\over 10^{14} \ {\rm g \
	cm^{-3}}}\right)\left({U_{o}\over 100 \ {\rm 
	keV}}\right)^{-1}\left({r_{o} \over 10 \ {\rm 
	fm}}\right)^{2}\left({\bar l \over 50 \ {\rm fm}}\right)^{-1}.
\end{equation}

Let the angular velocity {\em lag} between the superfluid and the
crust be $\omega \equiv \Omega_{s} - \Omega_{c}$.  The critical 
angular velocity difference above which the Magnus force prevents 
vortex pinning is $\omega_{c} \equiv F_{p}/r\rho_{s} \kappa \bar l$.  
The number of pinning sites involved in an unpinning event is 
determined by the value of $\tau$.  When $\tau > 1$, many pinning 
bonds must be broken for a vortex segment to unpin.  Exact expressions 
for the activation energy in this limit can be found in Link \&
Epstein (1991). 
In the limiting cases of $\omega << \omega_{c}$ and $\omega \simeq 
\omega_{c}$, $E_{a}$ is (eqs.  [B.12] and [5.1], Link \& Epstein 
1991),

\begin{equation}
E_{a} \simeq \left\{ \begin{array}{ll} 2.2 
U_{o}\sqrt\tau\left({\omega_{c}\over\omega}\right) & {\omega << 
\omega_{c}} \\
			5.1 U_{o}\sqrt\tau(1 - {\omega\over\omega_c})^{5/4} & {\omega 
			\simeq \omega_{c}}.
						\end{array}
						\right .
\end{equation}
In our simulations, we use the exact expressions for $E_{a}$ given in 
Link \& Epstein \shortcite{LE91}.

\subsection{Thermal Dynamics}

Changes in the local temperature affect the vortex creep rate and 
hence the rotation rate of the star.  A temperature enhancement 
generates a thermal wave which propagates through the star according 
to the thermal diffusion equation

\begin{equation}
	c_{v}{\partial T \over \partial t} = \nabla \cdot 
	(\kappa_{T}\nabla T) + H_{{\rm friction}} - \Lambda_{\nu}.
\label{Diff}
\end{equation}
Here $c_{v}$ is the specific heat and $\kappa_{T}$ is the 
thermal conductivity, both of which depend on density and temperature.  
The internal heating rate from friction between the superfluid and 
crust is $H_{{\rm friction}}$ and $\Lambda_{\nu}$ is the cooling rate 
through neutrino emission.  The heating rate due to superfluid 
friction \cite{Alp84,SL,Umeda,VRLE} is,

\begin{equation}
H_{{\rm friction}} = \int { dI_{s} \omega \vert\dot\Omega_{s}\vert}. 
\label{SFH}
\end{equation}

Relevant cooling mechanisms include neutrino cooling via the modified 
URCA process, neutron-neutron and neutron-proton bremsstrahlung in the 
core \cite{FM}, and electron bremsstrahlung \cite{I84} in the crust.  
The neutrino cooling rates are:

\begin{equation}
	\Lambda_{\nu}^{{\rm URCA}} = 1.8\times 
	10^{21}m_{n}^{*3}m_{p}^{*}\left(\frac{\rho} 
	{\rho_{o}}\right)^{2/3}T_{9}^{8} \ {\rm ergs \ cm^{-3} \ s^{-1}}
\end{equation}

\begin{equation}
	\Lambda_{\nu}^{{\rm nn}} = 4.4\times 
	10^{19}m_{n}^{*4}\left(\frac {\rho}{\rho_{o}}\right)^{1/3}T_{9}^{8} 
	\ {\rm ergs \ cm^{-3} \ s^{-1}} 
\end{equation}

\begin{equation}
	\Lambda_{\nu}^{{\rm np}} = 5.0\times 
	10^{19}m_{n}^{*2}m_{p}^{*2}\left(\frac 
	{\rho}{\rho_{o}}\right)^{2/3}T_{9}^{8} \ {\rm ergs \ cm^{-3} \ 
	s^{-1}}
\end{equation}

\begin{equation}
	\Lambda_{\nu}^{{\rm brem}} = 1.6\times 
	10^{20}\left(\frac{Z^{2}}{A}\right)\left(\frac{\rho}{\rho_{o}}\right)
	T_{9}^{6} \ {\rm ergs \ cm^{-3} \ s^{-1}},
\end{equation}
where $m_{n}^{*}$ is the ratio of the effective mass to the bare mass 
of the neutron, and similarly for the proton.  We take $m_{n}^{*} = 
m_{p}^{*} = 0.8$ in our calculations \cite{BKS}.  Our values for $A$ 
(the ion mass number) and $Z$ (the ion proton number) are from Lattimer 
\etal \shortcite{LPRL}, and $T_{9}$ is the internal temperature in 
units of $10^{9}$ K.

We take the surface of the star to cool through blackbody radiation,

\begin{equation}
	L_{bb} = 4 \pi \sigma R_{\infty}T_{s,\infty}^{4},
\label{bb}
\end{equation}
where $R_{\infty}$ and $T_{s,\infty}$ are the radius and surface 
temperature seen by a distant observer.  These quantities are related 
to their values at the surface through the redshift $e^{-\Phi} \equiv 
(1 - 2GM/Rc^{2})^{-1/2}$ as

\begin{equation}
 T_{s,\infty} = e^{\Phi}T_{s}, \qquad R_{\infty} = e^{-\Phi}R.	
\end{equation}

The specific heat of the star is due predominantly to degenerate 
electrons \cite{GS} with significant contributions from the ions at 
lower densities (Van Riper 1991; see Chong \& Cheng 1994 for 
corrections).  The thermal conductivity is a function of density and 
temperature.  We use the results of Itoh \etal \shortcite{I83,I84} and 
Mitake, Ichimaru \& Itoh \shortcite{MII} for a pure crust.

Impurities may arise in the crust as a result of the cooling history
of the star.  Early in the star's thermal evolution ($T \simeq
10^{10}$ K) lattice crystallization is expected to occur more
quickly than beta equilibration processes \cite{FlowRud,Jon99}.
Consequently, nuclei with different nuclear charge (impurities) from
the dominant nuclei are likely to be formed.  The concentration of
impurities lowers the mean-free-path of the electrons, reducing the
thermal conductivity in the crust.  The electron-impurity thermal
conductivity $\kappa_{T,Q}$ is (see e.g. Ziman 1972, eq.  7.92),

\begin{equation}
	\kappa_{T,Q} = \frac{\pi^{2}n_{e}k^{2}T}{3 m_{e}^{*}}\tau_{eQ},
\end{equation}
where $n_{e}$ is the density of electrons, $m_{e}^{*}$ is the 
effective mass of the electron, $k$ is the Boltzmann constant, and 
$\tau_{eQ}$ is the electron-impurity relaxation time.  For a high 
concentration of impurities, the impurity relaxation time can be 
approximated by the electron-ion relaxation time \cite{YU}.

\begin{equation}
	\tau_{ei} = \frac{p_{F}^{2}v_{F}}{4\pi 
	Z^{2}e^{4}n_{N}}\Lambda_{ei}^{-1},
\end{equation}
where $p_{F}$ and $v_{F}$ are the momentum and velocity of an electron 
at the Fermi surface and $\Lambda_{ei} = {\rm ln}[(2\pi 
Z/3)^{1/3}(1.5+3/\Gamma)^{1/2}] - 1$.  The ion density is $n_{N}$ and 
$\Gamma$ is the lattice order parameter.  To obtain a lower limit on 
the thermal conductivity for an impure crust we calculate the 
conductivity due to electron-ion scattering, treating the ions as if 
they were liquified (see discussion in Brown 2000).  We obtain the 
liquid-state thermal conductivity numerically, using the results of 
Itoh \etal (1983) and Mitake \etal (1984).  We also use the 
liquid-state neutrino emissivity \cite{Haensel} in this case.

\section{Models}
\label{MDG}

We consider the transfer of angular momentum from the superfluid to
the crust through two mechanisms: 1) a deposition of thermal energy
which liberates pinned vortex lines from their pinning barriers (a
thermal glitch) and 2) mechanical motion of vortices (a mechanical
glitch).  The first case would arise from the heat deposition
associated with a starquake.  The second case applies if crust
cracking occurs through direct vortex forces as a result of strong
pinning \cite{Rud91}, or through magnetic stresses arising from the
forcing of the field through the crust by core vortices \cite{Rud98}.
Mechanism 2 would also describe a catastrophic unpinning event
\cite{Cheng88,AP,MI}.  In both models, the vortices are allowed to
creep as described in Section \ref{RD}.  We initiate vortex motion in
both models at a density of $1.5 \times 10^{14}$ g cm$^{-3}$,
representative of the densest regions of the inner crust, where the
contribution to the moment of inertia of pinned superfluid is largest.
Since this density is near nuclear density, where the lattice is
expected to dissolve, our treatment also models core-driven glitches
when the vortex motion occurs near the crust-core boundary
\cite{Rud98}.

Sudden vortex motion generates heat.  For a glitch which conserves 
angular momentum, the heating associated with the vortex motion 
$E_{\rm motion}$ is determined by the change in rotational energy.  
Angular momentum conservation gives

\begin{equation}
	I_{c}\Delta\Omega_{c} + \int dI_{s}\Delta\Omega_{s} = 0.
\end{equation}
The heat liberated is

\begin{equation}
	E_{\rm motion} = \Delta\left[\frac{1}{2}I_{c}\Omega_{c}^{2} + 
	\frac{1}{2}\int dI_{s}\Omega_{s}^{2}\right] = \int {dI_{s} \omega 
	({\bf r}) \Delta\Omega_{s}({\bf r})}.
\label{VorH}
\end{equation}
Here $\omega({\bf r}$) is the angular velocity lag before the glitch, 
and $\Delta\Omega_{s}({\bf r})$ is the change in the superfluid 
angular velocity due to vortex motion.

If vortices unpin in a mechanical glitch, vortex drag will limit the 
spin-up timescale to $\sim$ 100 rotation periods \cite{EB92} or $\sim 
1$ minute in the slowest rotating pulsar that we consider.  If vortex 
motion occurs through motion of crust material with little unpinning 
of vortices, the spin-up timescale is approximately the distance the 
vortices move divided by the transverse sound speed in the crust, $t 
\lap \Delta r/c_{t} \simeq 10^{-4}$ s, a small fraction of a rotation 
period.  Since these timescales are shorter than any time-resolved 
glitch observations, we approximate the spin-up in a mechanical
glitch as an infinitely fast transfer of angular momentum from the
superfluid to the crust. This produces a step-like initial increase in
the spin rate of the crust.

\subsection{Geometry}

A complete treatment of the thermal and rotational dynamics described 
above would require a multi-dimensional analysis.  The energy 
deposition due to a starquake could occur in a localized region 
\cite{LFE,FLE}; the subsequent thermal diffusion probably lacks any 
simple symmetry, making the vortex dynamics complicated.  Moreover, 
the heat dissipated by a moving vortex depends on position along the 
line, further complicating the dynamics.  These effects make the 
coupled thermal and rotational dynamics a difficult two or 
three-dimensional problem.  We adopt a one-dimensional (radial) 
treatment as a first step in modeling thermal and rotational changes, 
as we now describe.

The vortices of the rotating superfluid would align themselves with the 
rotation axis of the solid if there were no pinning.  The number of 
vortex lines present in the superfluid is
\begin{equation}
	N = \frac{2 \pi R^{2} \Omega_{s}}{\kappa} \simeq \frac{10^{16}}{P},
\end{equation}
where $P$ is the spin period in seconds.  A bundle of $\Delta N$ 
vortex lines has $\Delta N$ times the circulation and therefore 
$(\Delta N)^{2}$ times the tension of a single line (eq.  
\ref{tension}) and effectively resists bending.  Consequently, a 
bundle of vortices tends to remain straight and aligned with the 
rotation axis even when forces vary along the bundle.  We therefore 
treat the vortex array as infinitely stiff over the dimensions of the 
crust and average the vortex creep velocity along a vortex.  In this 
approximation, the vortices are always aligned with the rotation axis 
of the crust.  We take the vortex distribution to be axi-symmetric, 
and follow changes in the superfluid rotation rate as a function of 
the distance from the rotation axis.

We follow thermal changes in the star by solving the thermal diffusion
equation (eq.  \ref{Diff}) with spherical symmetry.  Though a crude
approximation, this treatment of the thermal evolution captures the
essence of the dynamics while conserving energy.  We account for the
frictional heat generation, which has axi-symmetry, in the following
way.  The heat generated (eq. \ref{SFH}, supplemented by
eq. \ref{VorH} for a mechanical glitch) is integrated over each
cylindrical shell in the inner crust.  The total heat liberated is
then divided by the volume of the inner crust and is included as a
source term in the spherical treatment of the thermal diffusion.
Because the heat is distributed with spherical symmetry throughout the
crust, this approach underestimates the temperature increase in some
parts of the star.  We begin the stellar core at nuclear density and
treat it as isothermal.  The surface temperature is obtained by
matching the surface temperature to the internal temperature using the
prescription of Gudmundsson, Pethick \& Epstein \shortcite{GPE} with
equation \ref{bb} as a boundary condition on the heat flux.

\section{Simulations}

We model a $1.4 M_{\odot}$ neutron star using the equation of state of 
Friedman \& Pandharipande \shortcite{FP}.  The stellar radius is 10.5 
km and the central density is $\sim 1.0 \times 10^{15}$ g cm$^{-3}$.  
The moment of inertia of the effective crust (which includes the mass 
of the core plus the crust) is $I_{c} = 7.8 \times 
10^{44}$ g cm$^{2}$.  To compare with observed glitches, we begin our 
simulations using values of the spin parameters ($\Omega,\dot\Omega$) 
reported from observations.  We take the temperature of the Crab 
pulsar to be approximately half its observed upper limit \cite{CrabT}.  
We choose Vela's temperature in the middle of the observationally 
determined range \cite{VelaT}.  For the older pulsars PSR 1822-09 and 
PSR 0355+54, we take temperatures consistent with standard cooling 
models which include superfluid effects \cite{VR91}.  Temperatures and 
spin parameters for the pulsars in our study are listed in Table 
\ref{params}.

We solve for the initial lag $\omega({\mathbf r},t=0$) numerically 
using equation \ref{SFD} with the spin-down rate set equal to the 
observed value.  After a glitch is initiated, we solve the thermal 
diffusion equation to update the internal temperature.  The vortex 
creep velocity is then obtained using equation \ref{creep}.  Once the 
creep velocity is known, equation \ref{SFD} is solved for the 
superfluid rotation rate as a function of location in the star.  
Equation \ref{ANG} is used to obtain the angular velocity of the 
crust.  With the angular velocities of the superfluid and the crust 
now known, the lag is updated as the difference between the two and the 
entire process is repeated.  We evaluate the vortex stiffness $\tau$ 
at a density of $1.5\times10^{14} {\rm g \ cm}^{-3}$.  We choose a 
lattice spacing of $\bar l$= 50 fm which represents an average value 
for the inner crust (see Link \& Epstein 1991; Link, Epstein \& Baym 
1993).  We vary the pinning parameters $U_{0}$ and $r_{0}$ to obtain 
the best fit to the data, and take them to be constant throughout the 
crust.  At each time step the internal temperature is updated using 
the thermal diffusion equation (eq.  \ref{Diff}).

We initiate a thermal glitch with a spherically-symmetric deposition
of heat centered on a density of $1.5\times 10^{14} {\rm g \
cm}^{-3}$.  The shell has a gaussian distribution in radius with a
full width of 40 meters.  Density gradients in the inner crust are not
significant over this length scale, so the deposition is isolated to a
high density region.  A thermal glitch occurs in two phases. A quick
rise in the crust's spin rate, resulting from the initial energy
deposition, is followed by a slower rise as the thermal wave
dissipates. The angular velocity lag is reduced as a result of the
glitch, which in turn decreases the coupling between the superfluid
and the crust.  As a result, the external torque temporarily acts on a
smaller moment of inertia and the star spins down at a greater rate
than before the glitch.  In older (presumably colder) stars, which
have a lower specific heat, the energy deposition causes a larger
increase in temperature which generates a faster glitch than the same
energy deposition would in a younger star.  Our results depend on the
location of the energy deposition.  Less energy is required when the
deposition is at lower density to produce a similar spin response,
holding all other parameters constant.  A wider pulse requires more
total energy to generate the same spin response.

We initiate a mechanical glitch with an axi-symmetric impulsive change
in the superfluid angular velocity from its steady-state value
centered on a density of $1.5\times 10^{14} {\rm g \ cm}^{-3}$.  The
shell has a gaussian distribution in radius with a full width of 40
meters.  Sudden movement of the superfluid vortices generates heat, as
described in Section \ref{MDG} (see eq.  \ref{VorH}).  After the
initial angular velocity change, we solve for the thermal and
rotational response of the star.  A mechanical glitch occurs as a
step corresponding to the repositioning of vortex lines with respect
to their steady state locations.  Following the initial spin jump the
superfluid relaxes and eventually the lag is recovered.  The thermal
pulse associated with a mechanical glitch is orders of magnitude
smaller that that resulting from a thermal glitch.  This difference
occurs because the heat deposition required to mobilize pinned
vortices in a thermal glitch is much larger than the heat generated by
vortex motion in a mechanical glitch.  A larger change in the
superfluid angular velocity is required to produce the same spin
response if the vortex motion occurs at lower density, holding all
other parameters constant.  If the change in superfluid velocity
occurs through a wider shell, a smaller peak in the gaussian profile
is needed to give the same spin response.

The parameters which best fit the data, along with the $\chi^{2}$/dof, 
are listed in Tables \ref{fittherm} and \ref{fitmech} for the thermal 
and mechanical glitch models, respectively.  For comparison, the 
$\chi^{2}$/dof of the steady state observations is given (preglitch 
$\chi^{2}$/dof) as a measure of the inherent scatter in the data.  We 
report values for $U_{o}, r_{o}, \Delta E$ and $\bar\omega$, where 
$\Delta E$ is the energy deposited in a thermal glitch and 
$\bar\omega$ is the angular velocity difference between the superfluid 
and the crust, averaged over the superfluid moment of inertia.

\subsection{PSR 0531+21 (Crab Pulsar)}

\begin{figure}

\begin{center}

\epsfxsize=6.2in

\epsfysize=4.5in

\epsffile{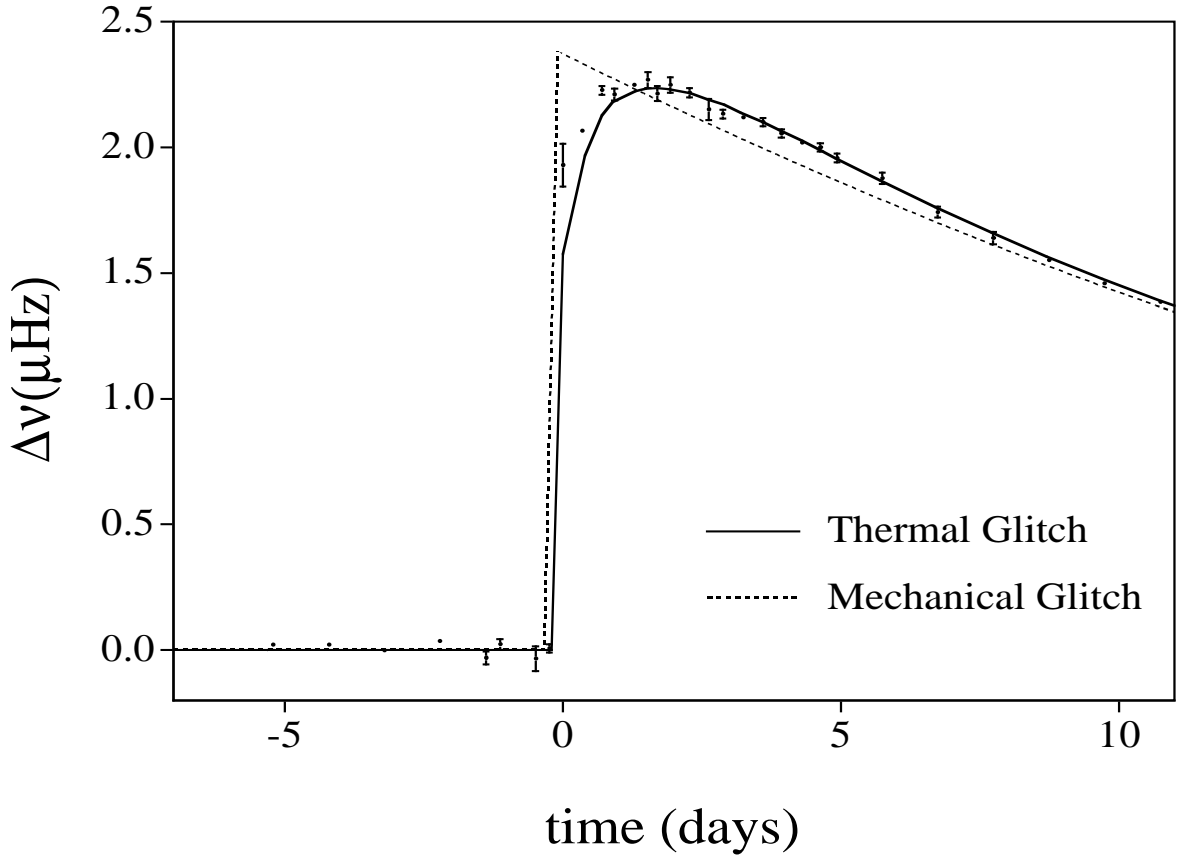}

\end{center}

\caption
{A thermal glitch ({\it solid line}) in the Crab pulsar after an 
energy deposition of $1.5 \times 10^{42}$ ergs.  A mechanical glitch 
({\it dashed line}), resulting from the sudden motion of superfluid 
vortex lines is also shown.  Data from the 1988 glitch are shown 
(Lyne, Smith \& Prichard 1992). The secular spin down has been
subtract. }
\label{CrabGlitch}
\end{figure}

\begin{figure}

\begin{center}

\epsfxsize=6.2in

\epsfysize=4.5in

\epsffile{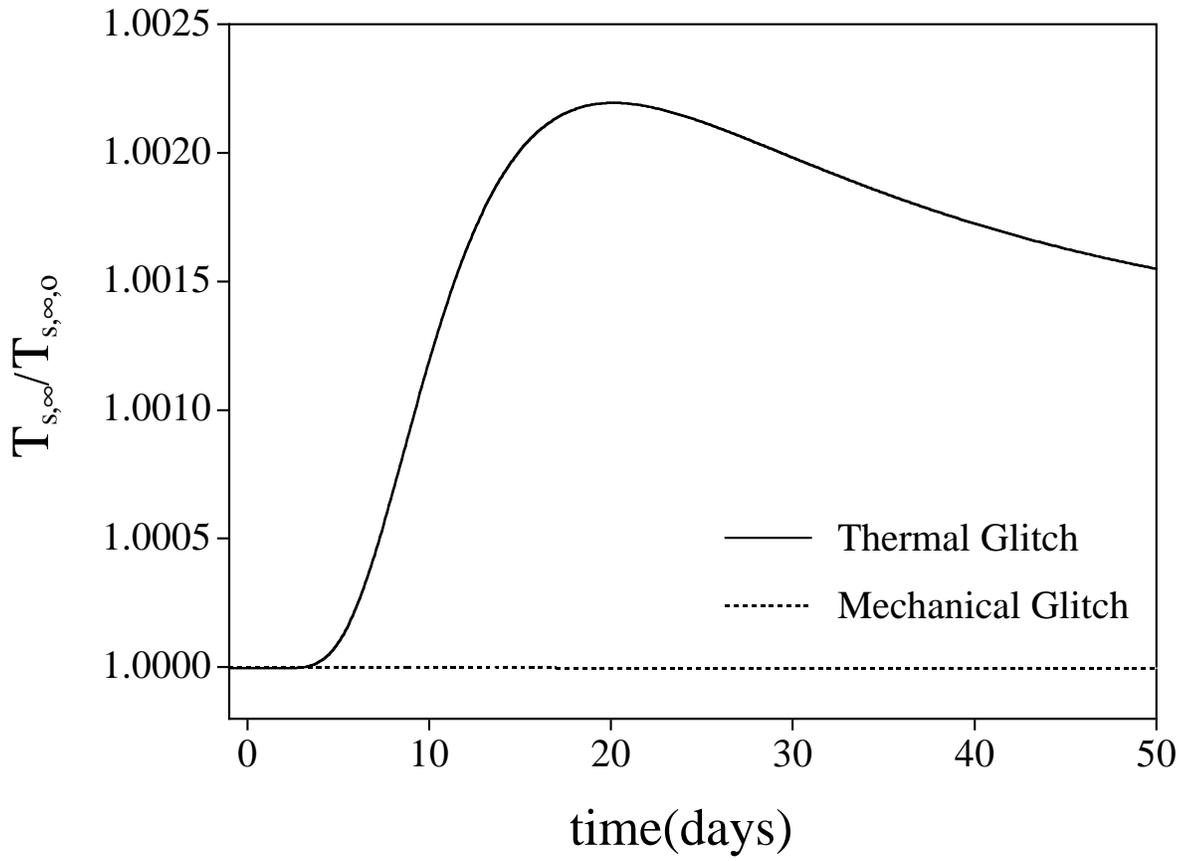}

\end{center}

\caption
{Surface temperature of the Crab pulsar after a thermal ({\it solid 
line}) and mechanical ({\it dashed line}) glitch in a pure crust 
model, in units of the initial temperature at infinity, 
$T_{s,\infty,o}$.}
\label{CrabSurface}
\end{figure}

Simulations of the 1988 glitch in the Crab pulsar \cite{Lyne} are
shown in Fig. \ref{CrabGlitch}.  The glitch is best modeled by a
thermal glitch with an energy deposition of $1.5 \times 10^{42}$
ergs (solid line).  The thermal pulse triggers transfer of angular
momentum to the crust over a timescale of minutes.  As the thermal
pulse dissipates, the remaining spin-up occurs over approximately one
day.  The glitch has a fractional increase in rotation rate of
$\Delta\nu/\nu \simeq 7 \times 10^{-8}$.  The mechanical glitch model
is unable to simulate the gradual rise of the Crab observations.
The quick spin up is followed by slow decay of the spin increase
(Fig.  \ref{CrabGlitch}, dashed line).  The thermal pulse associated
with the thermal glitch peaks at the stellar surface $\sim 20$ days
after glitch onset and shows a maximum temperature increase of $\sim
0.2\%$ (Fig.  \ref{CrabSurface}, solid line).  The surface temperature
increase associated with the mechanical glitch is of negligible
magnitude.

\subsection{PSR 0833-45 (Vela Pulsar)}

\begin{figure}

\begin{center}

\epsfxsize=6.2in

\epsfysize=4.5in

\epsffile{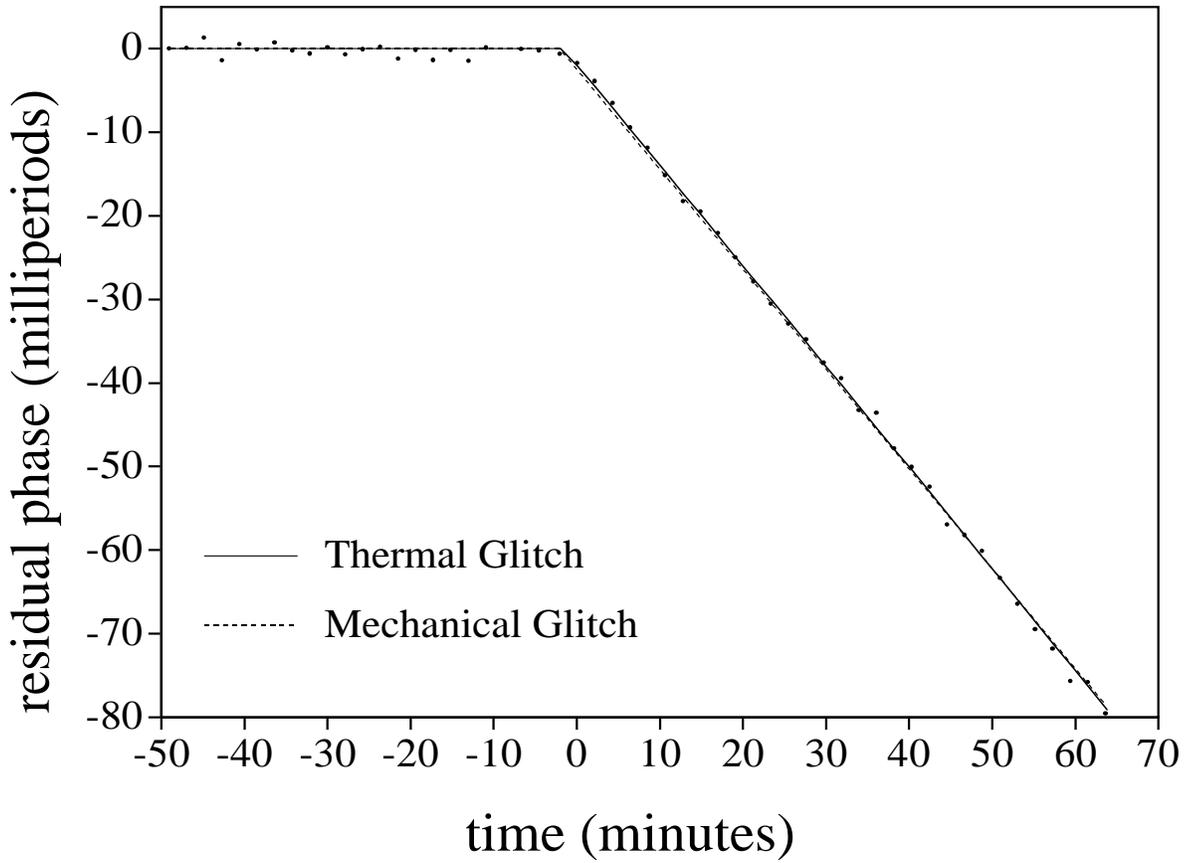}

\end{center}

\caption
{A thermal glitch in the Vela pulsar after an energy deposition of 
$6.5 \times 10^{42}$ ergs and a mechanical glitch resulting from the 
sudden motion of superfluid vortex lines.  The two models are nearly 
indistinguishable on this scale.  Data from the 1989 ``Christmas'' 
glitch are shown (McCulloch \etal 1990). The secular spin down has been
subtract. }
\label{VelaGlitch}
\end{figure}

\begin{figure}

\begin{center}

\epsfxsize=6.2in

\epsfysize=4.5in

\epsffile{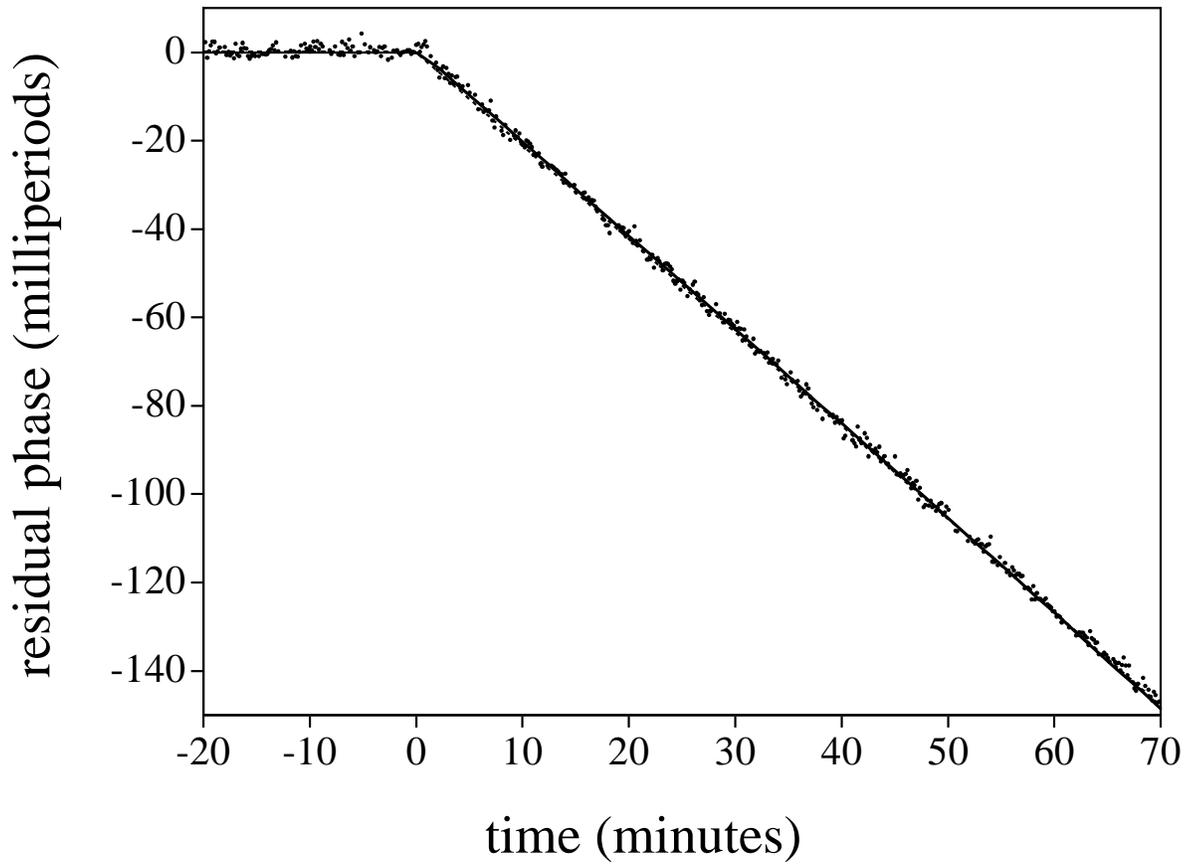}

\end{center}

\caption
{A thermal glitch in the Vela pulsar after an energy deposition of 
$7.5 \times 10^{42}$ ergs and a mechanical glitch resulting from the 
sudden motion of superfluid vortex lines.  The two models are 
indistinguishable on this scale.  Data from the January 2000 glitch 
are shown (Dodson \etal 2000). The secular spin down has been
subtract. }
\label{Vela2000Glitch}
\end{figure}

\begin{figure}

\begin{center}

\epsfxsize=6.2in

\epsfysize=4.5in

\epsffile{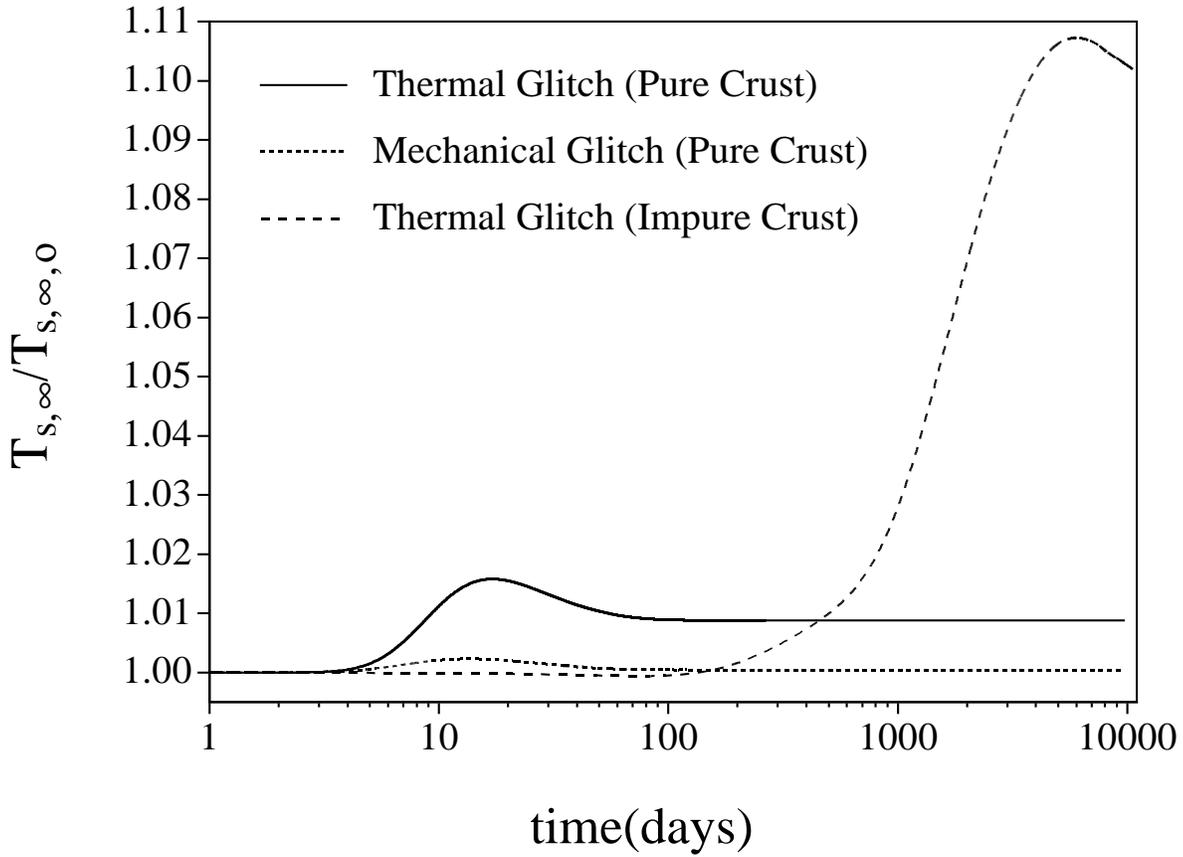}

\end{center}

\caption
{Surface temperature changes in the Vela pulsar after a thermal ({\it 
solid line}) and mechanical ({\it dashed line}) glitch in a pure crust 
model.  The {\it long-dash} line shows the surface temperature 
response for a highly impure crust after a thermal glitch.  These 
curves correspond to a glitch of magnitude, $\Delta \nu / \nu \simeq 3 
\times 10^{-6}$, as was observed in the Vela pulsar in January 2000.}
\label{Vela2000Surface}
\end{figure}

\begin{figure}

\begin{center}

\epsfxsize=6.2in

\epsfysize=4.5in

\epsffile{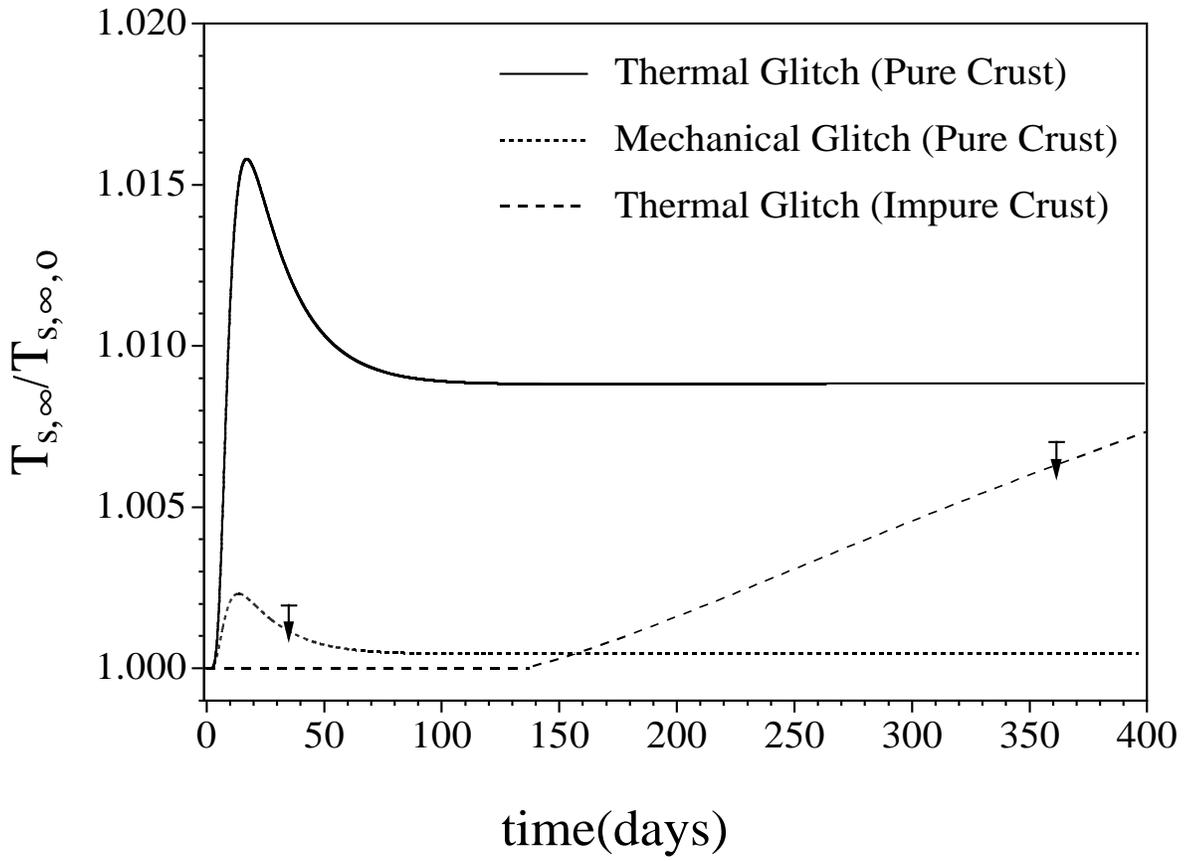}

\end{center}

\caption
{The first few months of Fig.  5.  Thermal observations 35 and 361 
days after the January 2000 glitch are indicated (Helfand \etal 2000; 
Pavlov \etal 2000; Pavlov, private communication).}
\label{Vela2000CloseSurface}
\end{figure}

The 1989 ``Christmas'' glitch ($\Delta\nu/\nu \simeq 2 \times
10^{-6}$) in the Vela pulsar \cite{Vela} is well modeled as either a
thermal or mechanical glitch (Fig.  \ref{VelaGlitch}); the simulations
are virtually indistinguishable.  The two models are also
indistinguishable in the January 2000 glitch \cite{DMC} in the Vela
pulsar (Fig.  \ref{Vela2000Glitch}).  The thermal pulse which reaches
the stellar surface differs significantly between the two mechanisms,
with a thermal glitch generating a pulse approximately 2 orders of
magnitude larger than that resulting from the mechanical glitch.  In
Figs. \ref{Vela2000Surface} and \ref{Vela2000CloseSurface} we show
the surface temperature increase following a thermal and mechanical
glitch for the January 2000 glitch in Vela.  CHANDRA observations
of thermal emission from the surface following the glitch in the Vela
pulsar on January 16, 2000 limit the temperature difference to $<
0.2\%$ 35 days after the glitch \cite{Helfand,Pavlov}, and $< 0.7\%$
361 days after the glitch (Pavlov, private communication). These
upper limits are marked in Fig. \ref{Vela2000CloseSurface}.  Lattice
impurities in the stellar crust would delay the arrival of the thermal
wave at the stellar surface.  The low thermal conductivity of an
impure crust also prevents the thermal wave from entering the
core, resulting in a larger increase in the surface temperature.  We
find that for a highly impure crust, the thermal pulse from a thermal
glitch peaks at the surface $\sim$ 16 years after the glitch, having a
magnitude of $\sim$ 10\% (Fig.  \ref{Vela2000Surface}).  Lowering the
thermal conductivity in this manner does not affect the spin-up
because vortex motion occurs over a relatively quick timescale during
the initial energy deposition.  The upper limits marked in Fig.
\ref{Vela2000CloseSurface} are inconsistent with a thermal glitch in a
pure crust but are consistent with a highly-impure crust.  Our
simulation of an impure crust gives an upper limit on the effects
that impurities could have. A smaller impurity fraction would decrease
the time at which the thermal pulse peaks at the surface.

\subsection{PSR 1822-09}

\begin{figure}

\begin{center}

\epsfxsize=6.2in

\epsfysize=4.5in

\epsffile{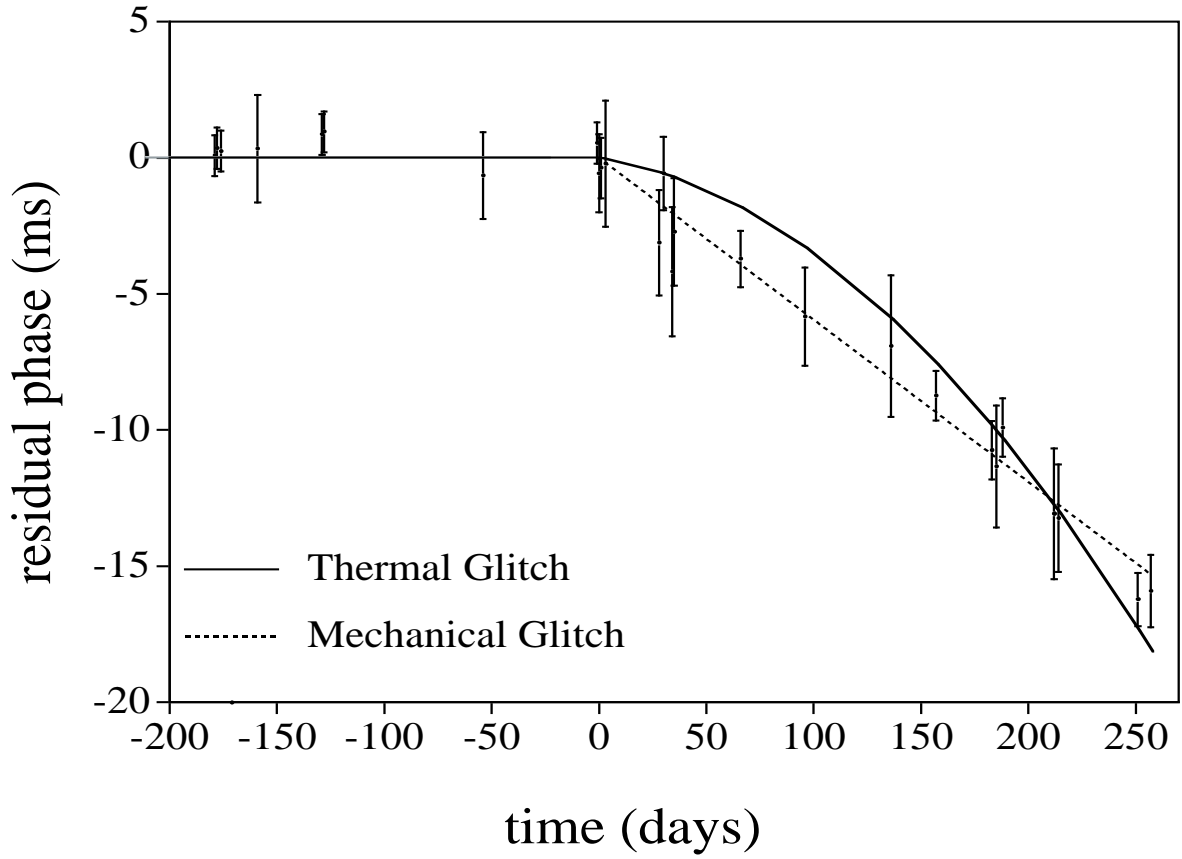}

\end{center}

\caption
{A thermal glitch ({\it solid line}) in PSR 1822-09 after an energy 
deposition of $4.3 \times 10^{41}$ ergs. A mechanical glitch ({\it 
dashed line}), resulting from the sudden motion of superfluid vortex 
lines, is also shown. Data from a 1994 glitch are shown 
(Shabanova 1998). The secular spin down has been
subtract. }
\label{1822Glitch}
\end{figure}

\begin{figure}

\begin{center}

\epsfxsize=6.2in

\epsfysize=4.5in

\epsffile{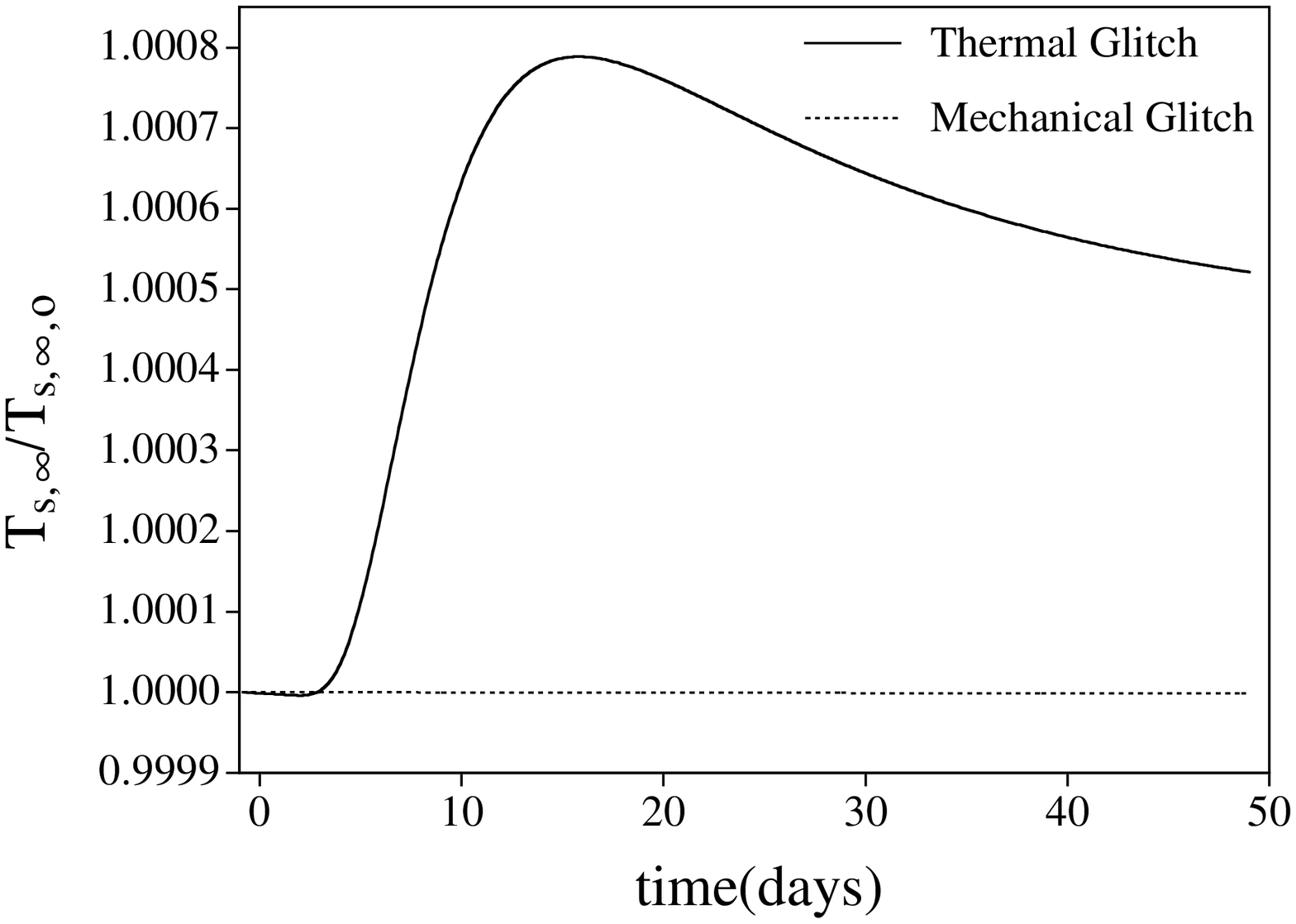}

\end{center}

\caption
{Surface temperature changes in PSR 1822-09 after a thermal ({\it 
solid line}) and mechanical ({\it dashed line}) glitch in a pure crust 
model.}
\label{1822Surface}
\end{figure}

Shabanova \shortcite{1822} observed an extremely small glitch, with a 
fractional change in spin rate of only $2.0 \times 10^{-10}$, in PSR 
1822-09 in September of 1994.  The thermal glitch simulation in Fig. 
\ref{1822Glitch} (solid line) turns over slowly as the thermal wave 
propagates through the star, consistent with the findings of Link \& 
Epstein \shortcite{LE}.  Although the mechanical glitch appears to the 
eye to be a better fit (Fig.  \ref{1822Glitch}, dashed line), both 
models are statistically consistent with the observations, having a 
$\chi^{2}$/dof within the scatter of the data.  The thermal pulse 
peaks at the surface of the star approximately 16 days after the onset 
of a thermal glitch (Fig.  \ref{1822Surface}) and has a maximum value 
of only 0.075\%.  There is a negligible increase in the surface 
temperature following a mechanical glitch.

\subsection{PSR 0355+54}

\begin{figure}

\begin{center}

\epsfxsize=6.2in

\epsfysize=4.5in

\epsffile{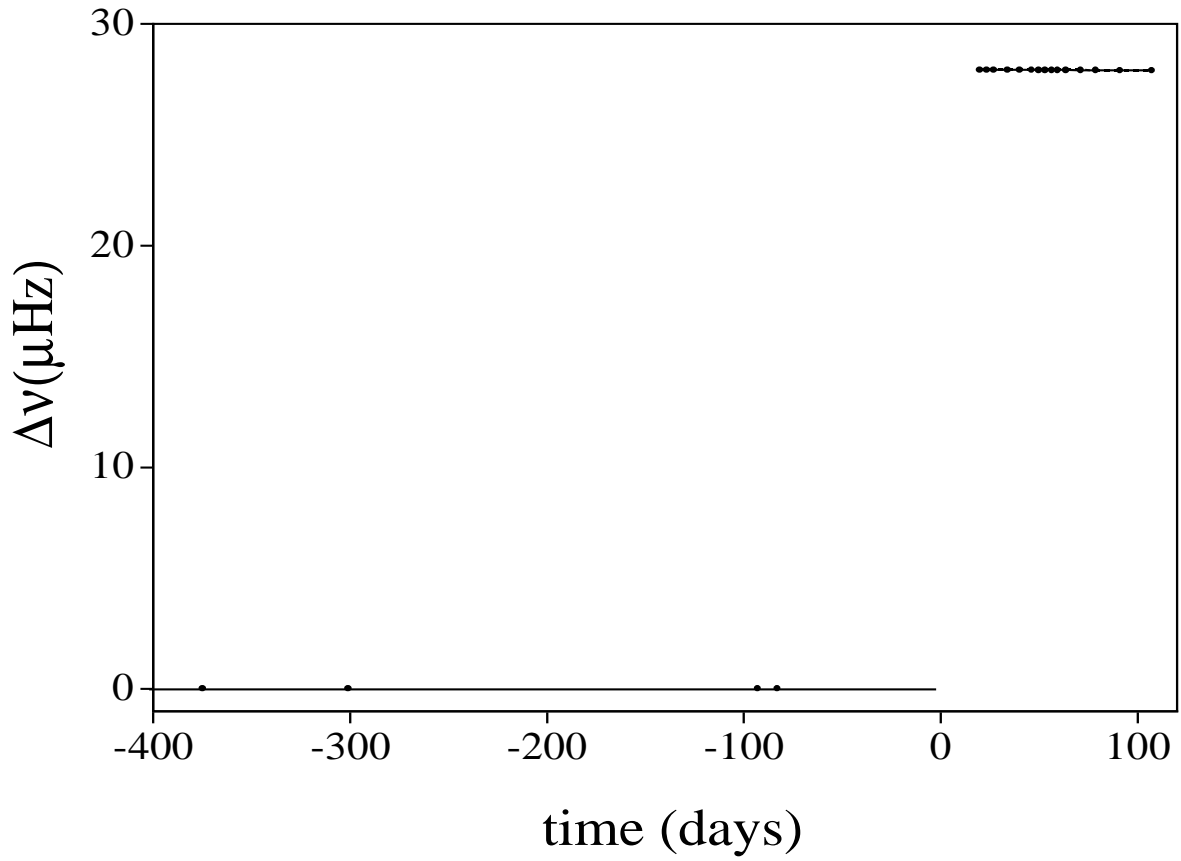}

\end{center}

\caption
{A thermal glitch in PSR 0355+54 after an energy deposition of $7.0 
\times 10^{42}$ ergs and a mechanical glitch resulting from the sudden 
motion of superfluid vortex lines.  The two models are 
indistinguishable on this scale.  Data from a 1986 glitch are shown 
(Shemar \& Lyne 1996). The secular spin down has been
subtract. }
\label{0355Glitch}
\end{figure}

\begin{figure}

\begin{center}

\epsfxsize=6.2in

\epsfysize=4.5in

\epsffile{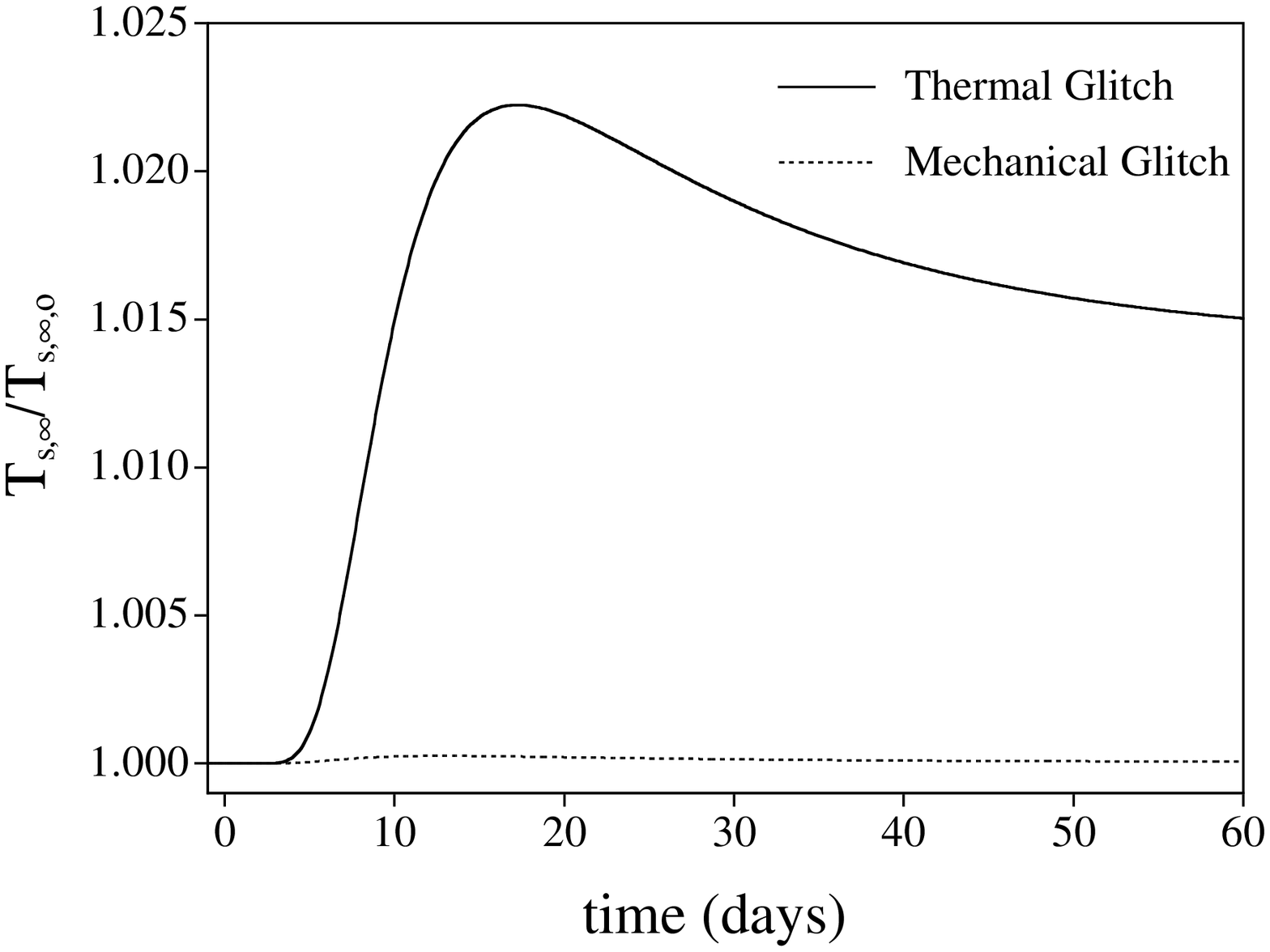}

\end{center}

\caption
{Surface temperature changes in PSR 0355+54 after a thermal ({\it 
solid line}) and mechanical ({\it dashed line}) glitch in a pure crust 
model.}
\label{0355Surface}
\end{figure}

The largest glitch observed in any pulsar occurred in PSR 0355+54 in 
March 1986 with a glitch magnitude of $\Delta\nu/\nu \simeq 4.4 \times 
10^{-6}$ \cite{0355}.  The glitch is well fit by either glitch model; 
the simulations are indistinguishable in Fig. \ref{0355Glitch}.  A 
thermal glitch generates a thermal pulse at the surface which peaks 
$\sim 17$ days after the glitch with a fractional temperature increase 
of $\sim 2\%$ (Fig.  \ref{0355Surface}, solid line).  A mechanical 
glitch generates a thermal pulse which peaks at a magnitude of $\sim 
0.02\%$, about 13 days after the initial spin up (Fig.  
\ref{0355Surface}, dashed line).

\section{Discussion}

This work has considered two different glitch scenarios involving the 
movement of superfluid vortex lines near nuclear density.  The thermal 
glitch model of Link \& Epstein \shortcite{LE} is consistent with all 
glitches modeled in this study.  The mechanical glitch model produces 
a spin jump that is too quick to explain the 1988 glitch in the Crab 
pulsar, but is consistent with glitch data in the other pulsars in our 
sample.  However, plastic flow of the crust following a mechanical 
glitch, an effect we have not accounted for, might provide a slow 
component to the spin-up in this younger, and presumably hotter star.

At present, timing measurements of glitching pulsars are too sparse to
distinguish between the predicted spin behavior of thermal and
mechanical glitches in most cases.  However, a distinguishing feature
between these models is the size of the surface temperature pulse
which accompanies the glitch.  The larger temperature increase
associated with a thermal glitch lowers the thermal conductivity,
slowing the pulse as it travels through the crust as compared to a
mechanical glitch.  In general, a thermal glitch in a pure crust
exhibits a thermal pulse at the stellar surface which is much larger,
and occurs a few days later, than a mechanical glitch. Impurities in
the stellar crust lower the thermal conductivity and could delay the
surface pulse by years.  A thermal glitch in an impure crust is
consistent with surface emission limits following the January 2000
glitch in the Vela pulsar \cite{Helfand,Pavlov}.  However, the
temperature simulation curve for an impure crust is still rising at
the time of the latest upper limit (see Fig. 6), and refinement
of this upper limit, as well as additional measurements, may conflict
with a thermal glitch in an impure crust as well.  Future glitch
observations coordinated with surface emission measurements will play
a key role in distinguishing between these two models. Further work is
needed to fully understand the effects of crust impurities on the
emerging thermal wave.

Other time-resolved observations of slow spin-ups in young pulsars,
similar to the Crab glitch of 1989, would support the thermal glitch
model.  A mechanical glitch does not occur slowly enough to
explain such behaviour, but is a viable model for glitches which
spin up the star quickly.  However, detection of an early thermal wave
of large amplitude could rule out the mechanical glitch model in a
pure crust.  It is possible that both thermal and mechanical
mechanisms are at work in generating pulsar glitches.

For both models we obtain effective pinning energies $U_{0}$ in the
range $\sim$ 10-500 keV. These values are less than calculated
pinning energies per nucleus \cite{Pizz} and may indicate the
presence of vortex tension effects which act to reduce the effective
pinning energy (Jones 1997;1998;1999; Link \& Cutler 2001).  The
relatively low values of $U_{0}$ we find for the Crab pulsar could
indicate that glitches originate in a different pinning region for
this object than in other pulsars.  The values we obtain for the range
of the pinning potential $r_{o}$ are consistent with existing
estimates \cite{Pizz,EB88}.  Larson \& Link (1999) obtained
constraints on the average angular velocity difference $\bar\omega$ 
assuming that superfluid friction accounts for the unexpectedly high
surface temperatures of some older pulsars.  That work also provided upper
limits on $\bar\omega$ in younger pulsars, by requiring that young
stars be stable against a thermal-rotational instability.  For stars
of the same age, our values of $\bar\omega$ are in agreement with
those results.

We thank K. Van Riper and K. Nomoto for providing subroutines to 
calculate the thermal conductivity and R.G. Dodson, A.G. Lyne, P.C. 
McCulloch, and T.V. Shabanova for providing pulsar glitch data.  We 
thank G.G. Pavlov for providing surface emission data from Vela and 
for useful discussions.  We also thank R.I. Epstein for valuable 
discussions.  This work was supported in part by NASA EPSCoR Grant No.  
291748 and NASA ATP No.  NAG 53688.  M.B.L thanks the Montana Space 
Grant Consortium for fellowship support during the time of this 
research.

\label{lastpage}

\clearpage

\begin{table*}

\begin{minipage}{102mm}

\caption{Physical parameters for the four pulsars used in our glitch 
simulations.}

\label{params}

\begin{tabular}{@{}lccccc}

\hline

	Pulsar & $t_{{\rm age}} \equiv \Omega_{c}/2\vert\dot\Omega_{c}\vert$ & 
	$\Omega_{c}$ & $\vert\dot\Omega_{c}\vert$ & $T_{{\rm int}}$ & 
	$T_{s,\infty}$ \\
	& (yr) & $({\rm rad \ s^{-1}})$ & ${\rm (rad \ s^{-2}})$ & (K) & (K) \\
	0531+21 & $1.2\times10^{3}$ & 189 & $2.4 \times 10^{-9}$ & 
	$8.6\times 10^{7}$ & $7.9 \times 10^{5}$ \\
	0833-45 & $1.1\times10^{4}$ & 70.4 & $1.0 \times 10^{-10}$ & $7.3 
	\times 10^{7}$ & $7.2 \times 10^{5}$ \\
	1822-09 & $2.3\times10^{5}$ & 8.2 & $5.6 \times 10^{-13}$ & $7.2 
	\times 10^{7}$ & $7.1 \times 10^{5}$ \\
	0355+54 & $1.2\times10^{6}$ & 40.2 & $1.1 \times 10^{-12}$ & $6.8 
	\times 10^{7}$ & $6.9 \times 10^{5}$ \\

\hline

\end{tabular}

\end{minipage}

\end{table*}

\begin{table*}

\begin{minipage}{100mm}

\caption{Internal parameters for the thermal glitch simulations.}

\label{fittherm}

\begin{tabular}{@{}lcccccc}

\hline

Pulsar & $U_{0}$ & $\overline\omega$  & $r_{0}$ & $\Delta$ E & 
	$\chi^{2}$/dof & $\chi^{2}$/dof \\
	       & (keV) &  ${\rm (rad \ s^{-1}})$  & (fm) & (ergs) & & (preglitch) \\
	0531+21 & 18 & 0.08  & 1.2 & $1.5\times 10^{42}$ & 10.5 & 11.2 \\
	0833-45 (1989) & 470 & 0.62  & 8.7 & $6.5\times 10^{42}$ & 0.55 & 0.55\\
	0833-45 (2000) & 554 & 0.78  & 8.1 & $7.5\times 10^{42}$ & 2.1 & 1.6\\
	1822-09 & 610 & 1.24 & 5.6 & $4.3\times 10^{41}$ & 1.2 & 1.0\\
	0355+54 & 62 & 0.08 & 8.0 & $7.0\times 10^{42}$ & 15.0 & 19.8\\ 

\hline

\end{tabular}

\end{minipage}

\end{table*}

\begin{table*}

\begin{minipage}{100mm}	

\caption{Internal parameters for the mechanical glitch simulations.}

\label{fitmech}

\begin{tabular}{@{}lccccc}

\hline
	Pulsar & $U_{0}$ & $\overline\omega$  & $r_{0}$ & $\chi^{2}$/dof &  
	$\chi^{2}$/dof \\
	       & (keV) &  ${\rm (rad \ s^{-1}})$ & (fm) & & (preglitch)\\
	0531+21 & 19 & 0.09 & 1.1 & 52.3 & 11.2 \\
	0833-45 (1989) & 220 & 0.57 & 4.2 & 0.84 & 0.55 \\
	0833-45 (2000) & 418 & 1.1 & 4.2 & 1.87 & 1.6 \\
	1822-09 & 520 & 1.18  & 4.9 & 0.52 & 1.0 \\
	0355+54 & 100 & 0.19  & 6.1 & 11.1 & 19.8 \\ 

\hline

\end{tabular}

\end{minipage}

\end{table*}

\clearpage

\end{document}